\documentclass{article}
\usepackage{enumitem}
\usepackage{spconf,amsmath,graphicx,amssymb}
\usepackage{color}
\PassOptionsToPackage{hyphens}{url}\usepackage{hyperref}
\newcommand{\vx}{\mathbf{x}}
\newcommand{\ve}{\mathbf{e}}
\newcommand{\vc}{\mathbf{c}}
\newcommand{\vw}{\mathbf{w}}
\newcommand{\ms}{\mathbf{S}}
\newcommand{\E}{\mathbb{E}}
\newcommand{\ie}{\textit{i.e.}, }

\def\x{{\mathbf x}}

\hypersetup{colorlinks=true}

\title{Generalized End-to-End Loss for Speaker Verification}
\name{Li Wan \qquad Quan Wang \qquad Alan Papir \qquad Ignacio Lopez Moreno\thanks{More information of this work can be found at: \url{https://google.github.io/speaker-id/publications/GE2E}}}

\address{Google Inc., USA\\[4pt] {
  \normalsize
  \{\href{mailto:liwan@google.com}{\nolinkurl{liwan}},
    \href{mailto:quanw@google.com}{\nolinkurl{quanw}},
    \href{mailto:papir@google.com}{\nolinkurl{papir}},
    \href{mailto:elnota@google.com}{\nolinkurl{elnota}}\}
    {\tt @google.com}
}}

\begin{document}
\ninept
\maketitle
\begin{abstract}
In this paper, we propose a new loss function called
generalized end-to-end (GE2E) loss, which makes the training of speaker
verification models more efficient than our previous tuple-based end-to-end~(TE2E) loss function.
Unlike TE2E, the GE2E loss function updates the network in a
way that emphasizes examples that are difficult to verify at each step of the
training process. Additionally, the GE2E loss does not require an initial stage
of example selection. With these properties, our model with the new loss function
decreases speaker verification EER by more than $10\% $, while reducing
the training time by $60\%$ at the same time.
We also introduce the MultiReader technique, which allows us to do domain
adaptation --- training a more accurate model that supports multiple
keywords (\ie ``OK Google" and ``Hey Google") as well as multiple dialects.

\end{abstract}
\section{Introduction}
\label{sec:intro}

\subsection{Background}
Speaker verification (SV) is the process of verifying whether an utterance
belongs to a specific speaker, based on that speaker's known utterances
(\ie enrollment utterances), with applications such as Voice Match \cite{multiuser,voicematch}.


Depending on the restrictions of the utterances used for enrollment and
verification, speaker verification models usually fall into one of two
categories: text-dependent speaker verification (TD-SV) and
text-independent speaker verification (TI-SV). In TD-SV,
the transcript of both enrollment and verification utterances is phonetially
constrained, while in TI-SV, there are no lexicon constraints on the
transcript of the enrollment or verification utterances, exposing a larger
variability of phonemes and utterance durations~\cite{kinnunen2010overview,bimbot2004tutorial}.
In this work, we focus on TI-SV and a particular subtask of TD-SV known as \emph{global password}
TD-SV, where the verification is based on a detected keyword, e.g. ``OK Google"~\cite{chen2014small,prabhavalkar2015automatic}

In previous studies, \textit{i-vector} based systems have been the
dominating approach for both TD-SV and TI-SV applications~\cite{dehak2011front}.
In recent years, more
efforts have been focusing on using neural networks for speaker
verification, while the most successful systems use
end-to-end training~\cite{variani2014deep, chen2015locally,Chao17,ZhangCZLG17,Seyed16}.
In such systems, the neural network output vectors are usually referred to as embedding vectors
(also known as \textit{d-vectors}). Similarly to as in the case of i-vectors, such embedding
can then be used to represent utterances in a fix dimensional space, in which other,
typically simpler, methods can be used to disambiguate among speakers.

\subsection{Tuple-Based End-to-End Loss}
In our previous work~\cite{heigold2016end},
we proposed the tuple-based end-to-end~(TE2E) model,
which simulates the two-stage process of runtime enrollment and verification during training.
In our experiments, the TE2E model combined with LSTM~\cite{hochreiter1997long} achieved the best performance
at the time.
For each training step, a tuple of one evaluation utterance $\vx_{j\sim}$ and $M$
enrollment utterances $\vx_{km}$ (for $m=1,\dots,M$) is fed into our LSTM network:
$\{\vx_{j\sim},(\vx_{k1},\dots, \vx_{kM})\}$,
where $\vx$ represents the features (log-mel-filterbank energies) from a fixed-length segment,
$j$ and $k$ represent the speakers of the utterances, and $j$ may or may not equal $k$.
The tuple includes a single utterance from speaker $j$ and $M$ different utterance from speaker $k$.
We call a tuple positive if $\x_{j\sim}$ and the $M$ enrollment utterances are
from the same speaker, \ie $j=k$, and negative otherwise.
We generate positive and negative tuples alternatively.

For each input tuple, we compute the L2 normalized response of the LSTM:
$\{\ve_{j\sim}, (\ve_{k1},\dots, \ve_{kM})\}$.
Here each $\ve$ is an embedding vector of fixed dimension that results from the
sequence-to-vector mapping defined by the LSTM. The centroid of
tuple $(\ve_{k1},\dots, \ve_{kM})$ represents the voiceprint built from $M$ utterances,
and is defined as follows:
\begin{equation}
  \label{eqn:centroid}
  \vc_k=\E_m [ \ve_{km} ]=\frac{1}{M}\sum_{m=1}^M \ve_{km}.
\end{equation}
The similarity is defined using the cosine similarity function:
\begin{equation}
\label{eqn:similarity_old}
  s=w\cdot \cos(\ve_{j\sim}, \vc_{k})+b,
\end{equation}
with learnable
$w$ and $b$.  The TE2E loss is finally defined as:
\begin{equation}
  L_{\mathrm{T}}(\ve_{j\sim}, \vc_k)=\delta(j,k)\Big(1- \sigma(s)\Big)+ \Big( 1-\delta(j,k) \Big) \sigma(s).
\end{equation}
Here $\sigma(x)=1/(1+e^{-x})$ is the standard sigmoid function and
$\delta(j,k)$ equals $1$ if $j=k$, otherwise equals to $0$.
The TE2E loss function encourages a larger value of $s$ when $k=j$, and a smaller
value of $s$ when $k\neq j$. Consider the update for both positive and
negative tuples --- this loss function
is very similar to the triplet loss in FaceNet~\cite{SchroffKP15}.

\subsection{Overview}
In this paper, we introduce a generalization of our TE2E architecture.
This new architecture constructs tuples from input sequences of
various lengths in a more efficient way, leading to a significant boost of
performance and training speed for both TD-SV and TI-SV. This paper is organized
as follows: In Sec. \ref{sec:ge2e} we give the definition of the GE2E loss;
Sec. \ref{sec:ge2etheory} is the theoretical
justification for why GE2E updates the model parameters more effectively;
Sec. \ref{sec:multireader} introduces a technique called ``MultiReader",
which enables us to train a single model that supports multiple keywords and
languages; Finally, we present our experimental results in Sec. \ref{sec:exp}.

\begin{figure*}[th]
  \centering
    \includegraphics[width=0.92\textwidth]{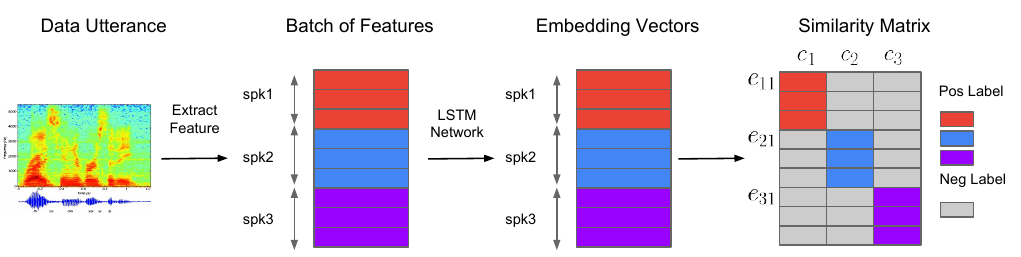}
  \caption{
    System overview. Different colors indicate utterances/embeddings from different speakers.
  }
  \label{fig:overview}
\end{figure*}

\section{Generalized End-to-End Model}
\label{sec:methods}
Generalized end-to-end (GE2E) training is based on processing a large number of
utterances at once, in the form of a batch that contains $N$ speakers,
and $M$ utterances from each speaker in average, as is depicted in Figure~\ref{fig:overview}.

\subsection{Training Method}
\label{sec:ge2e}
We fetch $N\times M$ utterances to build a batch.
These utterances are from $N$ different speakers, and each speaker has $M$ utterances.
Each feature vector $\vx_{ji}$ ($1 \leq j \leq N$ and $1 \leq i \leq M$)
represents the features extracted from speaker $j$ utterance $i$.

Similar to our previous work~\cite{heigold2016end}, we feed the features extracted from each utterance $\vx_{ji}$
into an LSTM network.
A linear layer is connected to the last LSTM layer as an additional
transformation of the last frame response of the network. We denote the
output of the entire neural network as $f(\vx_{ji};\vw)$ where
$\vw$ represents all parameters of the neural network (including both, LSTM layers and
the linear layer). The embedding vector (d-vector) is defined as the L2
normalization of the network output:
\begin{equation}
\label{eqn:lstm}
  \ve_{ji}=\frac{f(\vx_{ji};\vw)}{||f(\vx_{ji};\vw)||_2}.
\end{equation}
Here $\ve_{ji}$ represents the embedding vector of the $j$th speaker's $i$th
utterance. The centroid of the embedding vectors from the $j$th speaker
$[\ve_{j1},\dots, \ve_{jM}]$ is defined as $\vc_{j}$ via Equation~\ref{eqn:centroid}.

The similarity matrix $\ms_{ji,k}$ is defined as the scaled cosine similarities
between each embedding vector $\ve_{ji}$ to all centroids
$\vc_{k}$ ($1\leq j,k\leq N$, and $1\leq i\leq M$):
\begin{equation}
\label{eqn:similarity}
  \ms_{ji,k}=w\cdot \cos(\ve_{ji},\vc_k) + b ,
\end{equation}
where $w$ and $b$ are learnable parameters.
We constrain the weight to be positive $w>0$, because
we want the similarity to be larger when cosine similarity is larger.
The major difference between TE2E and GE2E is as follows:
\begin{itemize}
  \item TE2E's similarity (Equation~\ref{eqn:similarity_old}) is a scalar value
    that defines the similarity between embedding vector $\ve_{j\sim}$ and a single tuple centroid $\vc_k$.
  \item GE2E builds a similarity matrix (Equation~\ref{eqn:similarity}) that
    defines the similarities between each $\ve_{ji}$ and \textit{all} centroids $\vc_k$.
\end{itemize}
Figure~\ref{fig:overview} illustrates the whole process with features, embedding vectors, and
similarity scores from different speakers, represented by different colors.

\begin{figure}
  \centering
    \includegraphics[width=0.45\textwidth]{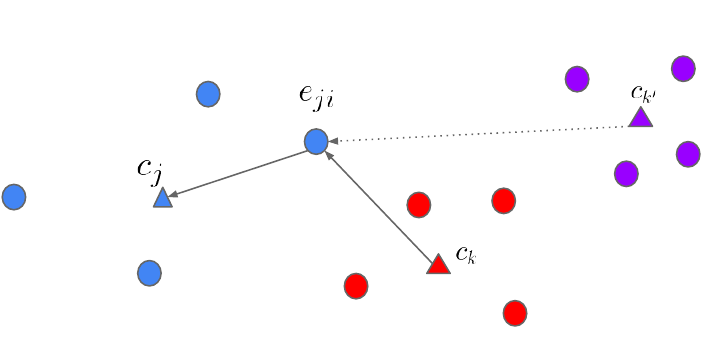}
  \caption{GE2E loss pushes the embedding towards the centroid of the true
  speaker, and away from the centroid of the most similar different speaker.}
  \label{fig:ge2e}
\end{figure}

During the training, we want the embedding of each utterance to be similar to the
centroid of all that speaker's embeddings, while at the same time,
far from other speakers' centroids. As shown in
the similarity matrix in Figure~\ref{fig:overview}, we want the similarity values of
colored areas to be large, and the values of gray areas to be small.
Figure~\ref{fig:ge2e} illustrates the same concept in a different way:
we want the blue embedding vector to be close to its own speaker's centroid (blue
triangle), and far from the others centroids (red and purple triangles), especially
the closest one (red triangle). Given
an embedding vector $\ve_{ji}$,
all centroids $\vc_{k}$, and the corresponding similarity matrix $\ms_{ji,k}$,
there are two ways to implement this concept:

{\bf Softmax} We put a softmax on $\ms_{ji,k}$ for $k=1,\dots,N$
	that makes the output equal to $1$ iff $k=j$, otherwise makes the output equal
	to $0$. Thus, the loss on each embedding vector $\ve_{ji}$ could be defined as:
	\begin{equation}
		\label{eqn:softmax}
		L(\ve_{ji})=-\ms_{ji,j}+\log\sum_{k=1}^N\exp(\ms_{ji,k}).
	\end{equation}
	This loss function means that we push each embedding vector close to its
	centroid and pull it away from {\it all} other centroids.

{\bf Contrast} The contrast loss is defined on positive pairs and most
	aggressive negative pairs, as:
	\begin{equation}
		\label{eqn:contrast}
		L(\ve_{ji})=1-\sigma (\ms_{ji,j}) + \max_{\substack{1\leq k\leq N\\k\neq j}} \sigma (\ms_{ji,k}),
	\end{equation}
	where $\sigma(x)=1/(1+e^{-x})$ is the sigmoid function.
	For every utterance, exactly two components are added to the loss:
	(1) A positive component, which is associated with a positive match
	between the embedding vector and its true speaker's voiceprint (centroid).
	(2) A \emph{hard} negative component, which is associated with a negative match
	between the embedding vector and the voiceprint (centroid) with the highest similarity among all
	false speakers.

In Figure~\ref{fig:ge2e}, the positive term corresponds to pushing $\ve_{ji}$ (blue circle) towards
$\vc_{j}$ (blue triangle).
The negative term corresponds to pulling $\ve_{ji}$ (blue circle)
away from $\vc_{k}$ (red triangle), because $\vc_k$ is more similar to $\ve_{ji}$
compared with $\vc_{k'}$.
Thus, contrast loss allows us to focus on difficult pairs of embedding vector and negative centroid.

In our experiments, we find both implementations of GE2E loss are useful: contrast loss
performs better for TD-SV, while softmax loss performs slightly better for TI-SV.

In addition, we observed that removing $\ve_{ji}$ when computing the centroid of the true speaker
makes training stable and helps avoid
trivial solutions.
So, while we still use Equation~\ref{eqn:centroid} when
calculating negative similarity (\ie $k\neq j$), we instead use
Equation~\ref{eqn:centroid2} when $k=j$:
\begin{eqnarray}
  \vc_{j}^{(-i)}&=&\frac{1}{M-1}\sum_{\substack{m=1\\m\neq i}}^{M}\ve_{jm},\label{eqn:centroid2} \\
  \ms_{ji,k} &=&
  \begin{cases}
    w\cdot \cos(\ve_{ji}, \vc_j^{(-i)})+b & \text{if} \quad k=j; \\
    w\cdot \cos(\ve_{ji}, \vc_k)+b & \text{otherwise}.
  \end{cases} \label{eqn:similarity2}
\end{eqnarray}

Combining Equations~\ref{eqn:lstm},~\ref{eqn:softmax},~\ref{eqn:contrast} and~\ref{eqn:similarity2},
the final GE2E loss $L_{G}$ is the sum of all losses over the similarity matrix ($1\leq j\leq N$, and $1\leq i\leq M$):
\begin{equation}
\label{eqn:ge2e}
  L_{G}(\vx;\vw)=L_{G}(\ms)=\sum_{j,i} L(\ve_{ji}).
\end{equation}

\subsection{Comparison between TE2E and GE2E}
\label{sec:ge2etheory}
Consider a single batch in GE2E loss update:
we have $N$ speakers, each with $M$ utterances. Each single step update will
push all $N\times M$ embedding vectors toward their own centroids, and pull them away the other centroids.

This mirrors what happens with all possible tuples in the TE2E loss
function~\cite{heigold2016end} for each $\vx_{ji}$. Assume we randomly choose
$P$ utterances from speaker $j$ when comparing speakers:
\begin{enumerate}
  \item Positive tuples: $\{\vx_{ji}, (\vx_{j,i_1},\dots,\vx_{j,i_P})\}$ for $1\leq i_p\leq M$ and $p=1,\dots, P$.
    There are $\binom{M}{P}$ such positive tuples.
  \item Negative tuples: $\{\vx_{ji}, (\vx_{k,i_1},\dots,\vx_{k,i_P})\}$ for $k\neq j$ and
    $1\leq i_p\leq M$ for $p=1,\ldots ,P$. For each $\vx_{ji}$,
    we have to compare with all other $N-1$ centroids,
    where each set of those $N-1$ comparisons contains $\binom{M}{P}$ tuples.
\end{enumerate}
Each positive tuple is balanced with a negative tuple,
thus the total number is the maximum number of positive and negative tuples times 2.
So, the total number of tuples in TE2E loss is:
\begin{equation}
  \label{eqn:num_updates}
  2\times \max\Big(\binom{M}{P}, (N-1)\binom{M}{P}\Big)\geq 2(N-1) .
\end{equation}
The lower bound of Equation~\ref{eqn:num_updates} occurs when $P=M$.
Thus, each update for $\vx_{ji}$ in our GE2E loss is identical to at least $2(N-1)$ steps
in our TE2E loss. The above analysis shows why GE2E updates models more
efficiently than TE2E, which is consistent with our empirical observations:
GE2E converges to a better model in shorter time (See Sec.~\ref{sec:exp} for details).

\subsection{Training with MultiReader}
\label{sec:multireader}
Consider the following case: we care about the model application in a domain
with a small dataset $D_1$. At the same time, we have a larger dataset $D_2$
in a similar, but not identical domain.
We want to train a single model that performs well on dataset $D_1$,
with the help from $D_2$:
\begin{equation}
  L(D_1,D_2;\vw)=\E_{x\in D_1}[L(\vx;\vw)] + \alpha \E_{x\in D_2}[L(\vx;\vw)] .
\end{equation}
This is similar to the regularization technique:
in normal regularization, we use $\alpha||\vw||_2^2$ to regularize the model. But
here, we use $\E_{x\in D_2}[L(\vx;\vw)]$ for regularization.
When dataset $D_1$ does not have sufficient data,
training the network on $D_1$ can lead to overfitting. Requiring the network
to also perform reasonably well on $D_2$ helps to regularize the network.

This can be generalized to combine $K$ different, possibly extremely unbalanced,
data sources: $D_1,\dots, D_K$. We assign a weight $\alpha_k$ to each data
source, indicating the importance of that data source.
During training, in each step we fetch one batch/tuple of
utterances from each data source, and compute the combined loss as:
$L(D_1,\cdots, D_K)=\sum_{k=1}^K \alpha_k \E_{\vx_k\in D_k}[L(\vx_k;\vw)]$,
where each $L(\vx_k;w)$ is the loss defined in Equation~\ref{eqn:ge2e}.

\section{Experiments}
\label{sec:exp}

In our experiments, the feature extraction process is the same as~\cite{prabhavalkar2015automatic}.
The audio signals are first transformed into frames of width 25ms and step 10ms.
Then we extract 40-dimension log-mel-filterbank energies as the features for
each frame. For TD-SV applications, the same features are used for both keyword
detection and speaker verification. The keyword detection system will only pass
the frames containing the keyword into the speaker verification system. These
frames form a fixed-length (usually 800ms) segment. For TI-SV applications, we
usually extract random fixed-length segments after Voice Activity Detection
(VAD), and use a sliding window approach for inference (discussed in Sec.~\ref{sec:tisid}) .

Our production system uses a 3-layer LSTM with projection~\cite{sak2014long}. The embedding vector
(d-vector) size is the same as the LSTM projection size. For TD-SV, we use $128$
hidden nodes and the projection size is $64$. For TI-SV, we use $768$ hidden
nodes with projection size $256$. When training the GE2E model, each batch contains $N=64$
speakers and $M=10$ utterances per speaker.
We train the network with SGD using initial learning rate $0.01$, and decrease it by half every 30M steps.
The L2-norm of gradient is clipped at $3$~\cite{Pascanu2012}, and the gradient scale for projection
node in LSTM is set to $0.5$. Regarding the scaling factor $(w,b)$ in loss function,
we also observed that a good initial value is $(w,b)=(10,-5)$, and
the smaller gradient scale of $0.01$ on them helped to smooth convergence.

\subsection{Text-Dependent Speaker Verification}
\label{sec:tdsid}
Though existing voice assistants usually only support a single keyword,
studies show that users prefer
that multiple keywords are supported at the same time. For multi-user on Google
Home, two keywords are supported simultaneously: ``OK Google" and ``Hey Google".

Enabling speaker verification on multiple keywords falls between TD-SV and TI-SV,
since the transcript is neither constrained to a single phrase, nor completely
unconstrained. We solve this problem using the MultiReader technique (Sec.~\ref{sec:multireader}).
MultiReader has a great advantage compared to simpler approaches,
e.g. directly mixing multiple data sources together:
It handles the case when different data sources are unbalanced in size.
In our case, we have two data sources for training:
1) An ``OK Google" training set from anonymized user queries with $\sim150$~M utterances and $\sim630$~K speakers;
2) A mixed ``OK/Hey Google" training set that is manually collected with $\sim1.2$~M utterances and $\sim18$~K speakers.
The first dataset is larger than the second by a factor of 125 in the number of utterances and 35 in the number of speakers.


For evaluation, we report the Equal Error Rate (EER) on four cases: enroll with either
keyword, and verify on either keyword. All evaluation datasets are manually collected from 665 speakers with an average of $4.5$ enrollment utterances and $10$ evaluation utterances per speaker.
The results are shown in Table \ref{tab:multireader}.
As we can see, MultiReader brings around 30\% relative improvement on all four
cases.

\begin{table}[t]
  \caption{MultiReader vs. directly mixing multiple data sources.}
  \label{tab:multireader}
  \begin{center}
  \scalebox{0.9}{
  \begin{tabular}{ l c c }
    \hline
    Test data & Mixed data & MultiReader \\
    (Enroll $\rightarrow$ Verify) & EER (\%) & EER (\%) \\ \hline \hline
    OK Google $\rightarrow$ OK Google & 1.16 & 0.82 \\
    OK Google $\rightarrow$ Hey Google & 4.47 & 2.99 \\
    Hey Google $\rightarrow$ OK Google & 3.30 & 2.30 \\
    Hey Google $\rightarrow$ Hey Google & 1.69 & 1.15 \\
    \hline
  \end{tabular}
  }
  \end{center}
\end{table}

We also performed more comprehensive evaluations in a larger dataset collected from $\sim 83$~K different speakers and environmental conditions, from both anonymized logs and manual collections. We use an average of $7.3$ enrollment utterances and $5$ evaluation utterances per speaker.
Table~\ref{tab:tdsv} summarizes average EER for different
loss functions trained with and without MultiReader setup. The baseline model is
a single layer LSTM with $512$ nodes and an embedding vector size of
$128$~\cite{heigold2016end}.
The second and third rows' model architecture is 3-layer LSTM.
Comparing the 2nd and 3rd rows, we see that GE2E is about $10\%$ better than TE2E.
Similar to Table \ref{tab:multireader}, here we also see that the model performs
significantly better with MultiReader. While not shown in the table, it is also
worth noting that the GE2E model took about $60\%$ less training time than TE2E.

\begin{table}[t]
  \caption{Text-dependent speaker verification EER.}
  \label{tab:tdsv}
  \begin{center}
  \scalebox{0.9}{
  \begin{tabular}{ l l l  c c}
    \hline
    Model         & Embed& Loss & Multi & Average \\
    Architecture  & Size &      & Reader & EER (\%) \\
    \hline \hline
    $(512,)$ \cite{heigold2016end} & $128$ & TE2E & No  & 3.30 \\
                                  &        &           & Yes & 2.78 \\
    \hline
    $(128,64)\times 3$ & $64$ & TE2E & No  & 3.55 \\
                       &     &  & Yes & 2.67 \\
    \hline
    $(128,64)\times 3$ & $64$ & GE2E & No  & 3.10\\
                       &     &  & Yes & 2.38 \\
    \hline
  \end{tabular}
  }
  \end{center}
\end{table}

\subsection{Text-Independent Speaker Verification}
\label{sec:tisid}

\begin{figure}[t]
  \centering
    \includegraphics[width=0.4\textwidth]{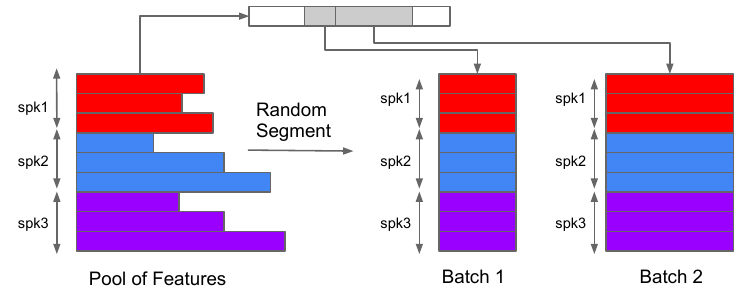}
  \caption{Batch construction process for training TI-SV models.}
  \label{fig:TI_train}
\end{figure}

For TI-SV training, we divide training utterances into smaller segments, which we
refer to as partial utterances. While we don't require all partial
utterances to be of the same length, all partial utterances in the same batch
must be of the same length. Thus, for each batch of data,
we randomly choose a time length $t$ within $[lb,ub]=[140,180]$ frames,
and enforce that all partial utterances in that batch are of length $t$
(as shown in Figure~\ref{fig:TI_train}).

\begin{figure}[t]
  \centering
    \includegraphics[width=0.45\textwidth]{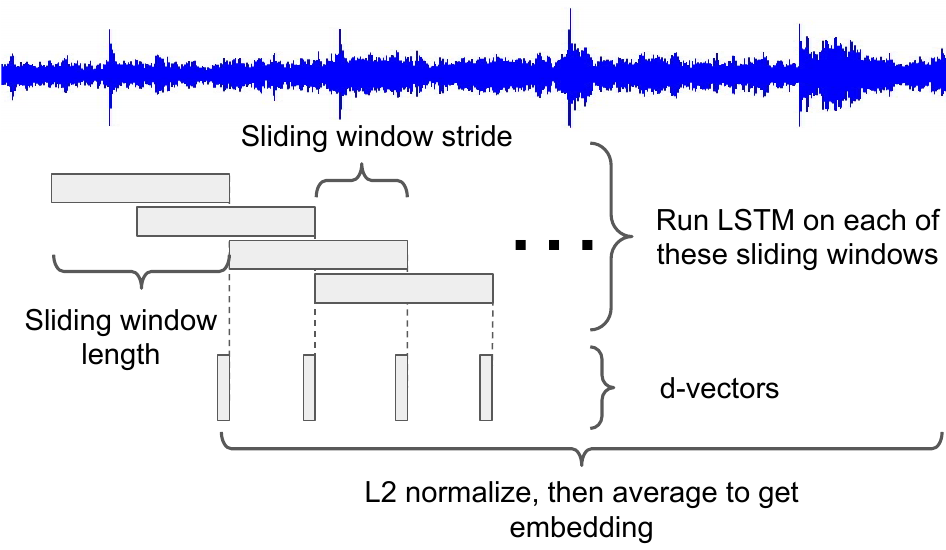}
  \caption{Sliding window used for TI-SV.}
  \label{fig:TI_inference}
\end{figure}

During inference time, for every utterance we apply a sliding window of fixed size
$(lb + ub)/2=160$ frames with $50\%$ overlap. We compute the d-vector for each window.
The final utterance-wise d-vector is generated by L2 normalizing the window-wise d-vectors,
then taking the element-wise averge (as shown in Figure~\ref{fig:TI_inference}).

Our TI-SV models are trained on around 36M utterances from 18K speakers, which are extracted from anonymized logs. For evaluation, we use an additional 1000 speakers with in average 6.3 enrollment utterances and 7.2 evaluation utterances per speaker.
Table~\ref{tab:tisv} shows the performance comparison between different training
loss functions.
The first column is a softmax that predicts the speaker label for all speakers in the training data.
The second column is a model trained with TE2E loss.
The third column is a model trained with GE2E loss.
As shown in the table, GE2E performs better than both softmax and TE2E.
The EER performance improvement is larger than  $10\%$.
In addition, we also observed
that GE2E training was about $3\times$ faster than the other loss functions.


\begin{table}[th]
  \caption{
    Text-independent speaker verification EER (\%).
  }
  \label{tab:tisv}
  \begin{center}
  \scalebox{0.9}{
  \begin{tabular}{ l c c c}
    \hline
    Softmax & TE2E~\cite{heigold2016end} & GE2E \\
    \hline \hline
    4.06 & 4.13 & 3.55 \\
    \hline
  \end{tabular}
  }
  \end{center}
\end{table}

\section{Conclusions}
\label{sec:conclusions}
In this paper, we proposed the generalized end-to-end (GE2E) loss function to train
speaker verification models more efficiently. Both theoretical and experimental results
verified the advantage of this novel loss function.
We also introduced the MultiReader technique to combine different data sources, enabling
our models to support multiple keywords and multiple languages. By combining
these two techniques, we produced more accurate speaker verification models.

%

\newpage
\bibliographystyle{IEEEbib}
\bibliography{refs}

\end{document}